\documentclass[aps,prl,twocolumn,showpacs,amsmath,amssymb]
{revtex4-1}
\usepackage{graphicx}
\usepackage{color}
\begin{document}
\title{Efficient Microcanonical Histogram Analysis and Application 
to Peptide Aggregation}
\author{Michael Bachmann}
\email{bachmann@smsyslab.org; https://www.smsyslab.org}
\affiliation{Soft Matter Systems Research Group, Center for Simulational
Physics, Department of Physics and Astronomy, The University of Georgia, 
Athens, GA 30602, USA}
\begin{abstract}
A novel approach designed to directly estimate microcanonical 
quantities from energy histograms is proposed, which enables the immediate 
systematic identification and classification of phase transitions in physical 
systems of any size by means of the recently introduced  generalized 
microcanonical inflection-point analysis method. The application to the 
aggregation problem of GNNQQNY heptapeptides, for which the entire transition 
sequence is revealed, shows the power of this promising method.
\end{abstract}
\maketitle 
In recent decades, computer simulations have become the inevitable tool for 
systematically approaching a better understanding of the physical behavior of 
complex systems. Statistical physics has provided the basis for the analysis 
of the vast amount of data generated in simulations of advanced models ever 
more precisely accommodating details of the physical systems. The development 
of efficient simulation techniques has kept pace with the demand by 
science and technology and the consistently increasing computational 
resources. 

For the important questions aiming at the cooperative behavior of 
microscopic agents such as atoms or molecules forming stable macroscopic 
phases, 
Monte Carlo and thermostated molecular dynamics simulation methods have 
proven to be particularly useful. Among the most widely used are 
generalized-ensemble Monte Carlo sampling methods such as parallel 
tempering~\cite{sw1,huku1,huku2,geyer1}, 
simulated tempering~\cite{marinari,lyubartsev}, multicanonical 
sampling~\cite{muca1,muca2,muca3,muca4,mbbook1}, and the Wang-Landau 
method~\cite{wl1,wl2}. 
These techniques were developed to address arguably the most difficult 
structure formation processes in many-body systems, including spin glasses 
and protein folding. The latter is a particularly interesting example, 
because by their very nature, proteins are finite systems on mesoscopic 
scale, yet exhibiting strong features of cooperativity known from phase 
transitions in macroscopic systems. Integrating finite, mesoscopic systems 
into the theory of phase transitions is a vital research problem.

The method proposed here can be readily employed to any set of 
energy histograms or energy probability distributions. The most obvious 
application is the 
combination of the individual histograms obtained in 
computer simulations at multiple temperatures, although reweighted histograms 
from 
generalized-ensemble simulations can also be used, provided additional 
weights such as multicanonical weight functions artificially introduced to 
enhance the simulation performance are divided out. 

For the following 
description of the method, we assume a set of $I$ canonical energy histograms 
$h_i(E)$, 
$i=1,2,\ldots,I$ is available. These histograms are the typical outcomes 
of parallel-tempering Monte Carlo simulations, where $I$ Metropolis 
Monte Carlo simulations run in parallel threads $i$. Exchanging replicas 
between the threads supports the decorrelation of data and, therefore, 
improves performance and data quality. This is particularly relevant in 
simulations across phase transition regions.

The canonical energy histogram obtained in the $i$th simulation thread at 
the canonical temperature $T_i^\mathrm{can}$ is related to the probability 
density distribution $p_i(E)$ at $T_i^\mathrm{can}$:
\begin{equation}
\label{eq:hist}
h_i(E)\sim p_i(E)=\frac{1}{Z(T_i^\mathrm{can})} 
g(E)e^{-E/k_\mathrm{B}T_i^\mathrm{can}},
\end{equation}
where $g(E)$ is the density of states, $Z(T_i^\mathrm{can})$ the canonical 
partition function at $T_i^\mathrm{can}$, and $k_\mathrm{B}$ the Boltzmann 
factor. As an absolute quantity, $Z$ is not accessible in typical 
importance-sampling Monte Carlo methods, but usually not of major interest 
anyway. The key quantity is the density of states as it is not only used for 
calculations of energetic averages and response quantities such as the 
specific heat, but it allows for the introduction of the microcanonical 
entropy
\begin{equation}
\label{eq:ent}
S(E)=k_\mathrm{B}\ln g(E).
\end{equation}
This is vital for the recently introduced 
generalized microcanonical inflection-point analysis method for the 
systematic identification and classification of phase transitions in 
physical systems of any size~\cite{qb1}, which has found traction in 
different fields of 
physics~\cite{sb1,rizzi1,pettini1,pettini2,dicairano1,ab1}. 

If the simulation was capable of covering the 
entire energy region of interest, an estimate of the density of states could 
be directly 
obtained from the histogram by dividing out the Boltzmann factor. Methods 
like parallel tempering that can easily be  
parallelized usually only provide fragmented information, though.

\begin{figure}
\centerline{\includegraphics[width=8.8cm]{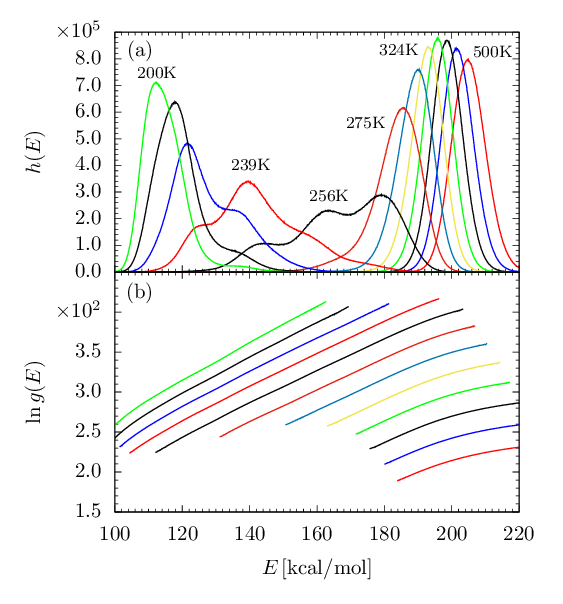}}
\caption{\label{fig:histent}%
(a) Raw histograms from parallel tempering simulations of GNNQQNY 
aggregation in 12 threads with temperatures in the range 
$T_i^\mathrm{can}\in[200\mathrm{K}, 500\mathrm{K}]$, $i=1,2,\ldots,12$. 
A few simulation temperatures are mentioned. (b) Entropy estimators 
$\mathrm{ln}\,g(E)$ obtained by naively reweighting the individual 
histograms.}
\end{figure}

Before discussing the reweighting method and microcanonical analysis from 
histogram data, a first look at results from actual parallel tempering 
simulations may be helpful. Figure~\ref{fig:histent}(a) shows the canonical 
energy histograms obtained in simulations of four 
identical heptapeptide chains at twelve different temperatures in the range 
$T_i^\mathrm{can}\in[200\mathrm{K},500\mathrm{K}]$ ($i=1,\dots,12$). The 
amino acid sequence of an 
individual chain is GNNQQNY. For these simulations, the 
internally developed Monte Carlo simulation package for proteins BONSAI 
(bio-organic nucleation and self-assembly at interfaces)
was employed. The protein model used for the simulations is based on an 
all-atom, 
implicit-solvent representation~\cite{irbaeck1,irbaeck2}. Each histogram 
contains the statistics of 
$10^8$ Monte Carlo sweeps, where a sweep is a sequence of updates including 
torsional rotations about dihedral angles of individual amino 
acids and rigid-body rotations and translations of entire chains.

The aggregation properties of GNNQQNY, which is a polar 
segment of the yeast protein Sup35, have been subject to numerous 
studies~\cite{eisenberg1,caflisch1,strodel1,obs1}. Individual 
chains have the tendency to form $\beta$-strands that can 
nucleate with other chains. Accumulating sufficient statistics 
for a detailed microcanonical analysis of the nucleation transition that 
lends insight into the aggregation mechanism across the energy region 
covering the entire transition sequence is virtually impossible. This can 
already be seen in Fig.~\ref{fig:histent}(b), where the individual estimates 
for the microcanonical entropy, obtained by simply dividing out the Boltzmann 
factor in the histograms, are plotted. Since the tails of those 
curves are 
hampered by low statistics in the histograms, the data quality significantly 
varies, even for a single fragment.

In order to obtain a reliable estimate for the microcanonical entropy (or the 
density of states), it is necessary to combine these fragmented estimators. 
This is not straightforward, because an absolute reference point is missing; 
the partition function cannot be estimated absolutely in these simulations as 
mentioned above. In fact, the relative shifts between the fragments shown in 
Fig.~\ref{fig:histent}(b) are not artificial. These are the actual numerical 
results based on the histogram data.

For an efficient estimation of the density of states, the multi-histogram 
reweighting method~\cite{sw2,wham} was already introduced decades ago. In 
this method, error 
weights are 
introduced to account for the low statistics in the tails of the canonical 
histograms. The error-weighted combination of histograms leads to a system 
of two equations that needs to be solved recursively. The result is an 
estimate for the density of states, which usually is then used in a 
subsequent canonical analysis, i.e., calculations of energetic quantities 
such as internal energy and heat capacity. This is a popular method that has 
proven its value countless times.

As more computing power became available throughout the years, the number of 
temperature threads in parallel tempering simulations increased. Large 
numbers of histograms can cause convergence problems in the recursive 
reweighting scheme, though. Also, directly estimating other microcanonical 
quantities that derive from the entropy is more beneficial for a 
microcanonical statistical analysis~\cite{qb1} than the estimate of the 
entropy itself.

For this reason, an error-reweighted histogram method without recursion is 
introduced in the 
following that also makes use of the same histogram data, but aims at a 
best-possible, direct estimate for the first derivative of the entropy with 
respect to energy, i.e., the microcanonical temperature:
\begin{equation}
\label{eq:beta}
\beta(E)=\frac{dS(E)}{dE}.
\end{equation}
It is important to note that the curvature features of the $\beta(E)$ 
estimate can already be used for the identification of all transitions 
the system experiences, independently of their order.

From Eq.~(\ref{eq:hist}), we conclude that the $i$th estimator for the 
entropy is given by $\hat{S}_i(E)=k_\mathrm{B}\ln 
h_i(E)+E/T_i^\mathrm{can}+c_i$, where $c_i$ is constant at 
$T_i^\mathrm{can}$. In the simulations, the energy space needs to be 
discretized. Assuming a uniform discretization, let $\Delta E$ be 
the energy difference between any two neighboring histogram bins centered at 
energies $E$ and $E-\Delta E$. Then, the $i$th estimator for the 
microcanonical inverse temperature can be written as:
\begin{equation}
\label{eq:betaest}
\hat{\beta}_i(E)=k_\mathrm{B}\left[\ln h_i(E)-
\ln h_i(E-\Delta E)\right]/\Delta E +1/T_i^\mathrm{can}.
\end{equation}
Following the similar simple assumption as in the original reweighting 
approach~\cite{sw2,wham} and in the multicanonical 
recursion~\cite{muca3,mbbook1} that in the simulations the 
energy bins are hit sufficiently rarely and that such hits are uncorrelated, 
the hit frequency follows a Poisson distribution. In this case, the variance 
of the histogram at the energy $E$ is identical 
to the average histogram value, $\sigma^2_h(E)=\langle h\rangle(E)$. 
Furthermore, it is assumed that $\langle h\rangle(E)\approx h(E)$. This allows 
us to estimate the error of the logarithm of $h$ from
\begin{equation}
\ln(h\pm\sigma_h)\approx\ln h\pm \frac{1}{\sqrt{h}}.
\end{equation}
Hence, $\sigma^2_{\ln h}=1/h(E)$. As variances of Poisson 
distributions are additive, the variance for the estimator $\hat{\beta}_i$ in 
Eq.~(\ref{eq:betaest}) is
\begin{equation}
\label{eq:varbeta}
\sigma^2_{\hat{\beta}_i}=\sigma^2_{\ln h_i}(E)+\sigma^2_{\ln h_i}(E-\Delta E)=
\frac{h_i(E)+h_i(E-\Delta E)}{h_i(E) h_i(E-\Delta E)}.
\end{equation}
Using this combined variance as an error weight for the relevance of 
the $i$th histogram in the estimation of $\beta$ from all histograms, we can 
combine all $I$ estimators $\hat{\beta}_i$ to determine the inverse 
microcanonical temperature as
\begin{equation}
\label{eq:optbeta}
\beta_\mathrm{rew}(E)=\frac{\sum\limits_{i=1}^I w_i(E) 
\hat{\beta}_i(E)}{\sum\limits_{i=1}^I w_i(E)},
\end{equation}
where the error weights are simply $w_i(E)=1/\sigma^2_{\hat{\beta}_i}$. 
This is obviously an easy-to-implement result and can in this form already be 
used for 
the analysis of microcanonical features that may signal phase transitions. 
For this purpose, it would be straightforward to extend this approach to 
derive similarly simple estimators for higher-order derivatives of the 
entropy, e.g., $\gamma(E)=d^2S(E)/dE^2=d\beta(E)/dE$, which can  be used to 
search for inflection points of higher-order transitions. It should also be 
noted, that the quality of the estimate of $\beta(E)$ allows for the recovery 
of the entropy $S(E)$ by simple numerical integration. From it the density 
of states $g(E)$ can be extracted and used for the canonical 
analysis of energy-dependent averages and response functions such as the heat 
capacity.

Instead of extending the above method to determine higher-order 
derivatives of $S(E)$, we follow a 
different approach here and further improve the estimator~(\ref{eq:optbeta}) 
by systematically reducing its numerical error by B\'ezier 
smoothing~\cite{mbbook1}. The 
result is an \emph{analytic function} in $E$. Hence, calculating the 
derivatives of 
the estimate for $\beta(E)$ is straightforward, providing also 
analytic estimates for the higher-order derivatives of 
the entropy. 

The B\'ezier method~\cite{mbbook1,bezier1,riesenfeld1} has long been used in 
computer graphics and design 
engineering to create smooth curves and surfaces, but it has not yet become a 
generic tool in systematically improving data quality in scientific 
applications. Given 
a discrete set of $N$ points $\mathbf{P}_n$ with $n=0,1,\ldots,N$, a 
continuous 
B\'ezier curve $\mathbf{B}(t)$, where $t\in [0,1]$ is the curve 
parameter, is constructed by
\begin{equation}
\label{eq:bez}
\mathbf{B}(t)=\sum\limits_{n=0}^N\mathcal{B}_n^{(N)}(t)\,\mathbf{P}_n.
\end{equation}
Here, 
\begin{equation}
\mathcal{B}_n^{(N)}(t)=\left(\begin{array}{cc}N\\ n\end{array}\right)
(1-t)^{N-n} t^n
\end{equation}
are the Bernstein polynomials. Since, for given $t$, these polynomials are 
identical to the binomial distribution and therefore already normalized, 
Eq.~(\ref{eq:bez}) can be interpreted as a curve construction method, where 
each point $\mathbf{P}_n$ contributes to the point $\mathbf{B}(t)$ of the 
B\'ezier curve with probability $\mathcal{B}_n^{(N)}(t)$. Alternatively, the 
position vectors
$\mathbf{P}_n$ can be imagined as forces pulling the smooth curve 
toward their 
position. It is obvious that this method may be extraordinarily beneficial for 
the smoothing of numerical data affected by numerical errors (or 
random noise). If a 
large number of (ideally) uncorrelated points are scattered around in the 
vicinity of $\mathbf{B}(t)$, the numerical error at $t$ can be 
reduced by this robust procedure. If $N$ is sufficiently large, the 
properties of the B\'ezier curve do not sensitively depend on it. 

In this form, the method works for one-dimensional curves in an embedding 
space of any dimension. Here, we are only interested in the two-dimensional 
space, where $\mathbf{P}_n=(E_n,\beta_\mathrm{rew}(E_n))$. If $n$ is the 
index of the histogram bin centered around $E_n$ in discretized energy space, 
Eq.~(\ref{eq:bez}) can simply be rewritten as~\cite{mbbook1}
\begin{eqnarray}
\label{eq:betabez}
&&\beta_\mathrm{bez}(E)=\\ \nonumber
&&\hspace*{5mm}\sum\limits_{n=0}^N
\left(\begin{array}{cc} N\\ n\end{array}\right) 
\left(\frac{E_N-E}{E_N-E_0}\right)^{N-n}
\left(\frac{E-E_0}{E_N-E_0}\right)^n 
\beta_\mathrm{rew}(E_n),
\end{eqnarray}
where the $\beta_\mathrm{rew}$ values at the discrete energies $E_n$ 
obtained by the reweighting method~(\ref{eq:optbeta}) are now used as control 
points for the generation of the B\'ezier curve. The range of the discrete 
energies is $[E_0,E_1,\dots,E_n,\ldots,E_N]$, where $E_0$ and $E_N$ are 
suitably chosen limits. Since the tails of the histograms close to the lowest 
and highest simulation temperatures may suffer from insufficient statistics, 
the boundaries of the energy range could be adjusted, but in effect the 
impact 
of those fringes on the results in the most interesting energy regions, e.g., 
near transition points, is negligible.
\begin{figure*}
\centerline{\includegraphics[width=16.6cm]{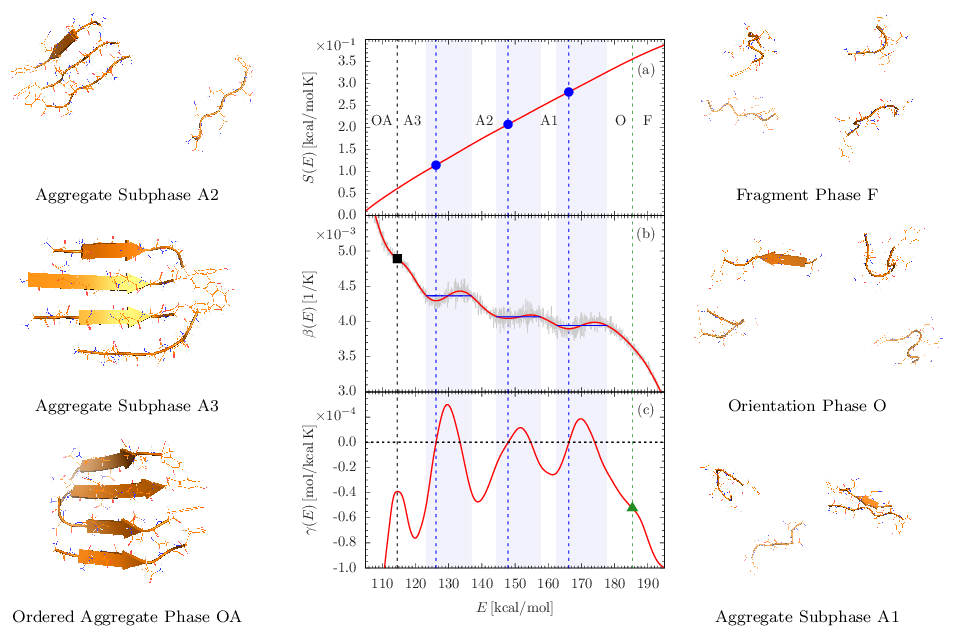}}
\caption{\label{fig:micro}%
(a) Microcanonical entropy $S(E)$ of four GNNQQNY 
chains. (b) Two estimates of the microcanonical inverse temperatures: 
$\beta_\mathrm{rew}(E)$ (grey) obtained by histogram reweighting and the 
corresponding B\'ezier curve $\beta_\mathrm{bez}(E)$ (red). (c) Second 
derivative of 
the entropy $\gamma(E)=d\beta(E)/dE=d^2S(E)/dE^2$ obtained by analytic 
differentiation of the B\'ezier curve $\beta_\mathrm{bez}(E)$ shown in (b). 
Circles in (a) and the square in (b) mark least-sensitive inflection 
points of these curves, indicating independent first- and second-order 
transitions, 
respectively. Maxwell constructions for the first-order transitions are 
shown as 
horizontal blue lines in (b) and blue shaded areas emphasize 
coexistence regimes; their widths correspond to the respective latent heats.
The triangle in (c) is located at the inflection-point of a 
dependent third-order transition in the disordered phase. Vertical dashed 
lines separate the individual phases. Representative conformations in the 
individual phases are also shown in atomic and cartoon representations. 
Arrows indicate that values of successive backbone dihedral angles enable the 
formation of $\beta$-strands in those sections of the chain.}
\end{figure*}

Since $\beta_\mathrm{bez}(E)$ is an analytic 
function, derivatives of 
$\beta_\mathrm{bez}(E)$ are easily obtained by differentiating 
Eq.~(\ref{eq:betabez}). 
The result can also be readily implemented; numerical differentiation is
not necessary. Higher-order analytic derivatives can be obtained accordingly. 
The maximum number of analytic derivatives is obviously only limited by the 
number of data points $\beta_\mathrm{rew}(E_n)$ with $n=0,1,\dots,N$.

The results of the application of the reweighting method and
B\'ezier smoothing to the GNNQQNY aggregation problem are 
shown in 
Fig.~\ref{fig:micro}. The direct estimate of the microcanonical inverse 
temperature obtained by the reweighting method, $\beta_\mathrm{rew}(E)$, is 
shown as the gray curve in Fig.~\ref{fig:micro}(b). This is a remarkable 
result. Despite the still visible numerical fluctuations, the characteristic 
backbending features indicating first-order transitions can uniquely be 
identified. The red curve is the continuous B\'ezier reconstruction 
$\beta_\mathrm{bez}(E)$ (with $N=1137$, $E_0=103.35\, 
\mathrm{kcal/mol}$, 
and $E_N=216.95\, \mathrm{kcal/mol}$) and enhances these features. The 
derivative of the 
B\'ezier curve, $\gamma(E)=d\beta_\mathrm{bez}(E)/dE$, is shown in 
Fig.~\ref{fig:micro}(c). The microcanonical entropy $S(E)$, plotted in 
Fig.~\ref{fig:micro}(a), is obtained as a byproduct by integrating 
$\beta_\mathrm{bez}(E)$.

The identification and classification of phase transitions and the 
description of the transition sequence in this system is 
straightforward with the precise information provided by the generalized 
microcanonical inflection-point analysis 
method~\cite{qb1}. Characteristic features of conformations 
in the energy regions associated with the structural phases such as hydrogen 
bonds, hydrophobic contacts between side chains, inter-chain 
contacts, helix and sheet contents, and chain orientations helped reveal all 
aspects of the 
transition sequence. In the fragment phase F, the individual chains are 
separate and do not exhibit structural features. This changes as the system 
undergoes a dependent third-order transition into the orientation phase O as 
indicated by the 
triangle in $\gamma(E)$ in Fig.~\ref{fig:micro}(c) at about 280~K. This is an 
important step in the sequence, because individual chains stretch out and 
even form oriented $\beta$-strands, which ultimately enables the inter-chain 
formation of hydrogen bonds once clusters start forming. 

It is worth noting that, in contrast to the 
typical independent 
transitions, a dependent transition cannot exist without an independent 
transition it is associated with. In this case, the corresponding independent 
transition is the closest first-order transition from O into the aggregate 
subphase A1, which initiates the 
aggregation sequence. If present, dependent transitions can 
only be found in the less ordered phase and therefore serve as a precursor 
for 
the major transition to occur upon reducing energy (or temperature).

Inflection points in $S(E)$ indicate the 
independent first-order transitions accompanying the aggregation process 
from A1 (single cluster of two chains) over A2 (one cluster with three 
chains or two clusters with two chains) to A3 (complete cluster). These are 
marked by circles in Fig.~\ref{fig:micro}(a). The 
identification is done best by looking for minima in $\beta(E)$, so $S(E)$ is 
not even needed for the actual identification of this type of transition. 
These findings are consistent with results  
from earlier studies of generic, coarse-grained polymer and 
protein models~\cite{jbj1,kb1}. As clusters of chains form, translational 
entropy is lost, which explains the effective entropic suppression at the 
corresponding transition energies. Not only do these results confirm the 
reality of these features, they also emphasize the capability of 
coarse-grained models to qualitatively reveal these subphase transitions 
accompanying the overall nucleation transition. If the system would contain 
more chains, the number of such first-order transitions increases in 
accordance with the combinatorial number of possible clusters. Eventually, 
toward the thermodynamic limit, the expectation is that the individual 
transitions merge into a single first-order nucleation phase 
transition~\cite{kb1}. 

This sequence of first-order-like features in nucleation processes is 
very common and has already been observed in the formation of finite atomic 
clusters long time ago~\cite{wales1,wales2}. Effectively, the cooperative 
effects of energy and entropy reduction create a barrier in the energy 
landscape $F(E,T_\mathrm{agg})=-k_\mathrm{B}T_\mathrm{agg}\ln\, 
Z_\mathrm{res}(E,T_\mathrm{agg})$. The individual histograms recorded at 
temperatures near the (canonical) aggregation temperature $T_\mathrm{agg}$, 
shown in Fig.~\ref{fig:histent}(a), can serve as estimators of the 
restricted partition function $Z_\mathrm{res}(E,T_\mathrm{agg})$. Energy 
barriers for each nucleation step originate from the respective bimodal 
shapes of the distributions in those energy regions, where two phases 
coexist~\cite{kb1}.  

The aggregate in A3 is mostly stabilized by hydrophobic interaction of 
the Tyrosine (Y) side chains at the C termini, but at the expense of 
hydrogen bonds along the strands. Eventually, an independent second-order 
transition into the ordered aggregate phase OA is found near the lower 
energy limit in the aggregated phase. The 
square in Fig.~\ref{fig:micro}(b) marks the corresponding inflection point in 
the inverse temperature curve and corresponds to the negative peak in 
$\gamma(E)$, which makes it easy to identify this transition. The 
transition can be attributed to the proper alignment of the strands by 
optimizing the number of hydrogen bonds for the formation of a 
well-organized $\beta$-sheet.   

To summarize, in this paper a novel microcanonical histogram reweighting 
method was introduced for the direct estimation of 
microcanonical quantities such as the microcanonical entropy, inverse 
temperature, and higher-order derivatives from basic energy histogram data 
typically obtained in standard computer simulations. In the context of the 
previously introduced generalized inflection-point analysis 
method~\cite{qb1}, curvature features of these quantities can then be used for 
the unique and systematic identification and classification of phase 
transitions in any physical system of any size. The power of these methods is 
demonstrated in the application to peptide aggregation. For a system of 
multiple GNNQQNY heptapeptides, the entire transition sequence ultimately 
leading to the formation of $\beta$-sheet assemblies could be identified. It 
will be exciting to compare transition hierarchies for different systems in 
future applications of the method introduced here.

The author thanks the Georgia Advanced Computing Resource
Center at the University of Georgia for providing computational
resources.

\end{document}